\documentstyle[aps,prb,twocolumn]{revtex}
\begin{document}

\begin{center}
{\Large Localization and entanglement of two interacting electrons in a 
quantum-dot molecule}

\bigskip
P.\ I.\ Tamborenea$^{\ast}$ and H.\ Metiu \\
Center for Quantized Electronic Structures (QUEST)
and Department of Chemistry \\
University of California \\
Santa Barbara, CA 93106-9510

\end{center}

\bigskip\noindent
{\it The localization of two interacting electrons in a coupled-quantum-dots 
semiconductor structure is demonstrated through numerical calculations of 
the time evolution of the two-electron wave function including the Coulomb 
interaction between the electrons.
The transition from the ground state to a localized state is induced by an 
external, time-dependent, uniform electric field.  
It is found that while an appropriate constant field can localize 
both electrons in one of the wells, oscillatory fields can induce 
roughly equal probabilities for both electrons to be localized 
in either well, generating an interesting type of localized and
entangled state.
We also show that shifting the field suddenly to an appropriate 
constant value can maintain in time both types of localization.}


\bigskip\noindent

Coherent control of quantum systems is at the heart of promising 
disciplines like femtochemistry and quantum information processing.
A basic operation of quantum control, namely, the localization 
of a single electron in coupled quantum wells, has been extensively
studied in the last decade.
In two early publications, conditions to maintain existing 
localization with an AC field,\cite{gro_dit_jun_han} and to create and
maintain localization with a semi-infinite AC field\cite{bav_met}
were identified.
Thereafter, localization
in two-level systems,
multilevel systems,
induced by ultrashort laser pulses, 
in dissipative environments, 
in molecular systems,
in trapped Bose-Einstein condensates,
by means of circularly polarized fields,
with bichromatic fields,
including the effect of Coulomb charging energy,
and with quantized electromagnetic fields,
has been studied.\cite{localization}

When two or more interacting particles are present the possibility 
of entanglement of the many-body wave function arises.
Entanglement is an essential ingredient in any scheme of quantum information
processing like quantum cryptography and quantum computation, and therefore 
it is a problem of great current interest to find or design systems where 
entanglement can be 
manipulated.\cite{steane,los_suk,qui_joh,ima_aws_bur,bre_gis_tit_zbi}

In the case of a single-electron, two-level system consisting of 
the lowest symmetric and antisymmetric states of the double well potential, 
perfect localization can be achieved with a strong periodic electric 
field that causes the two Floquet quasi-energies to be 
degenerate.\cite{two_level}
This scheme is not applicable to the two-electron system, because it 
requires a pair of states whose superposition results in a localized state, 
a condition that is not met in this system.
An alternative and trivial way of inducing localization in a two-electron 
double-well system is to adiabatically tilt the potential with a slowly 
increasing electric field.
Trying to localize the electrons on a short time scale is more
difficult: switching the field on rapidly excites higher electronic
states causing a coherent motion from one well to another.
In this Letter we examine this regime of very fast localization.

We investigate the localization and entanglement 
of two interacting electrons in a system of coupled quantum dots, induced 
by spatially uniform electric fields with a simple time dependence.
Systems of coupled quantum dots, sometimes referred to as
quantum-dot molecules or artificial molecules, have been actively 
investigated in the last five years, both experimentally and 
theoretically.\cite{coupled_dots_recent,tam-met,coupled_dots}
Our main findings are as follows: 
(i) A constant electric field can bring
the two electrons from their ground state (highly delocalized)
to a state of high degree of localization in one of the dots,
at a certain time;
(ii) At that time, a step in the field to another constant value
can maintain the localization essentially indefinitely;
(iii) An oscillatory field produces a different type of localization,
where both electrons are likely to be found together in either dot
with roughly equal probabilities.
This type of localization is a purely many-body phenomenon,
and arises due to the Coulomb interaction between the electrons,
which causes entanglement of the two-body wave function;
(iv) The entangled/localized states can also be maintained
in time by changing from the oscillatory field to an appropriate
constant field.
In all cases, localization takes place in a time scale of a few
picoseconds.

Our system is a quasi one-dimensional, double-quantum-dot
structure with two electrons in it.
The transversal size of the dots is taken to be $L=50 \, \mbox{\AA}$, 
and the double-well potential in the longitudinal direction,
$V(z)$, is shown in Fig.~1(a).
The energies associated with the transverse dimensions are, 
due to the narrow lateral confinement, 
high compared to those of the longitudinal motion.
Therefore, the lateral degrees of freedom do not participate in the 
dynamics and the two-electron wave function can be written as
(we discuss below the spin part of the wave function)
\begin{equation}
\Psi({\mathbf r}_1,{\mathbf r}_2,t)=\phi(x_1) \phi(y_1) \phi(x_2) 
\phi(y_2) \Phi(z_1,z_2,t).
\end{equation}
where $\phi(x)=\sqrt{2/L} \sin (\pi x/L)$.
The time-dependent Schr\"odinger equation becomes
\begin{eqnarray}
i\hbar \frac{\partial \Phi}{\partial t} &=&
    [-\frac{\hbar^2}{2m^{\ast}}
     \left(\frac{\partial^2}{\partial z_1^2}+\frac{\partial^2}{\partial z_1^2}
     \right) 
      +  V(z_1)+V(z_2) \nonumber \\
     &+& V_{1D}(|z_1-z_2|) -e(z_1+z_2) E(t) ] \Phi,
\label{eq:sch}
\end{eqnarray}
where $E(t)$ is an external time-dependent electric field, and $m^{\ast}$ 
is the effective mass.
$V_{1D}$ is the Coulomb interaction given by
\begin{eqnarray}
V_{1D}(|z_1-z_2|) &=& \int_0^L dx_1 dy_1 dx_2 dy_2 \nonumber \\
& & \frac{e^2 \phi^2(x_1) \phi^2(y_1) \phi^2(x_2) \phi^2(y_2)}
     {\epsilon |{\mathbf r}_1-{\mathbf r}_2|}.
\label{eq:V1d}
\end{eqnarray}
We use the effective mass $m^{\ast}$ and dielectric constant $\epsilon$ 
of GaAs.

In all our calculations the ground state is the initial state.
The spin part of the ground state of the two interacting electrons
is the singlet state,\cite{ash-mer} and correspondingly 
(the fermionic wave function is antisymmetric)
the spatial part of the wave function is symmetric under particle exchange.
We calculate the spatial part of the ground state by numerical diagonalization 
of the energy eigenvalue problem of the interacting electrons.
A plot of the ground state is shown in Fig.~1(b) as a function
of the $z$-coordinates of the two electrons.
Since the Hamiltonian of the system including the external electric
field is spin independent, the spin wave function is the singlet
at all times.
Therefore, the evolving spatial wave function remains symmetric
(under particle exchange) at all times.
We emphasize that the absence of triplet states in the expansion of
the wave function (Eq.~(\ref{eq:CI}) below) is {\em not} an approximation,
but a consequence of our choice of initial state and the lack of spin 
dependence in the Hamiltonian.

\begin{figure}
 \vbox to 8cm {\vss\hbox to 6cm
 {\hss\
   {\includegraphics{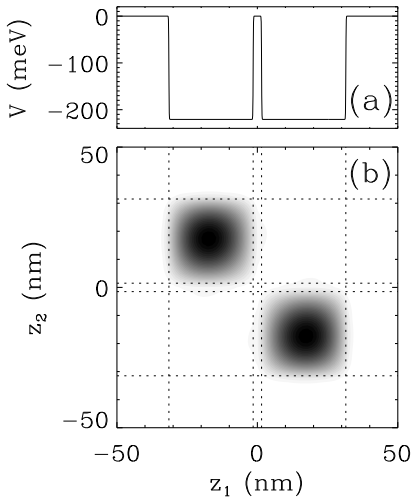}
   }
  \hss}
 }
\caption{
(a) Confining double-well potential in the longitudinal
direction of the coupled quantum dot structure;
(b) Two-electron ground state versus the longitudinal coordinates
of the two electrons.
Notice that the ground state is delocalized, ie. both electrons occupy
different dots.
}
\end{figure}

We calculate the evolution of the two-electron wave function 
by using the configuration interaction (CI) method,\cite{tam-met,sza_ost}
in which one expands the wave function as
\begin{equation}
\Phi(z_1,z_2,t)=\sum_{i,j} c_{ij}(t)
           [\varphi_i(z_1) \varphi_j(z_2) + \varphi_j(z_1) \varphi_i(z_2)],
\label{eq:CI}
\end{equation}
where $\varphi_i(z)$ are eigenstates of $P_z^2/2m^{\ast}+V(z)$.
The sum runs over $1 \leq i \leq j$.
We take $j=1,\ldots,N$, where $N=12$ is the number of bound states of 
the potential $V(z)$ shown in Fig.~1(a).
The two-particle basis set of symmetric states has $N(N+1)/2=78$
states.
We find that this basis set is large enough to achieve convergence.
To obtain the time evolution, the expansion (\ref{eq:CI}) is substituted 
in the Schr\"odinger equation (\ref{eq:sch}) and the coefficients 
$c_{ij}(t)$ are calculated numerically with the fourth-order 
Runge-Kutta method.

In order to describe the localization of the electrons, we introduce the
probabilities
\begin{equation}
P_{RL}(t) = 2 \, \int_R dz_1 \int_L dz_2 |\Phi(z_1,z_2,t)|^2,
\end{equation}
that one electron is in the right and the other one is in the left dot, 
\begin{equation}
P_{RR}(t) = \int_R dz_1 \int_R dz_2 |\Phi(z_1,z_2,t)|^2,
\label{eq:PRR}
\end{equation}
that both electrons are in the right dot, and
$P_{LL}(t)$, that both electrons are in the left dot.
Since the probability of ionization is kept small at all times
$P_{RL}(t)+P_{RR}(t)+P_{LL}(t) \approx 1$.

For the ground state, shown in Fig.~1(b), $P_{RL} \approx 0.9988$
and $P_{RR}=P_{LL} \approx 0.0006$.
That is to say that in the ground state the two electrons have
a very small probability to be found in the same well, 
which is expected due to their Coulomb repulsion.
In the rest of this Letter we explore the question of whether an 
external electric field (time dependent but spatially uniform) 
can induce localization of the two electrons on a very fast
(picosecond) time scale.

We start the search for localization with the simplest case, a 
constant electric field, $E(t)=E_0$.
Before $t=0$ the system is in the ground state, and at $t=0$ the 
field $E_0$ is switched-on suddenly.
We compute the evolution of the two-electron wave function as 
well as the probability $P_{RL}(t)$ during a simulation interval
of 9~ps.
For each field $E_0$ we find the lowest value of $P_{RL}$ that 
occurs within that time interval, and we plot the result as a function
of $E_0$ in Fig.~2(a).
The minimal $P_{RL}$ shows a few peaks but for only one value of $E_0$ 
($E_0=-5.18$~kV/cm) does it drop below 0.1.
To take a closer look at the peak with strongest localization, we 
calculate the three probabilities $P_{RL}, P_{RR}$, and $P_{LL}$ 
for that field, and plot them versus time in Fig.~3(a).
In this case, $P_{RL}$ and $P_{RR}$ show an oscillatory behavior, while
$P_{LL}$ remains negligible at all times.
Physically, the two electrons oscillate between a state in which
they are highly localized in the right well and another state
which is completely delocalized.

In Fig.~3(b) we illustrate two operations of control of the wave function 
that can be performed with piecewise-constant electric fields.
First, the thick-line curve of $P_{RL}(t)$ in Fig.~3(b)
(produced by the electric field shown by a thick line in Fig.~3(c))
indicates that the low value of $P_{RL}$ obtained earlier
(at $t\approx 3 \, \mbox{ps}$ in Fig.~3(a)) can be maintained
permanently by switching the field to another special value.
The field value that locks $P_{RL}$ at its minimum was
found through a systematic search.
The second operation consists of un-locking the localization and
resuming oscillations similar to those produced by the initial
constant field.
The timing of this second switching is found to be immaterial.
The new field value is not arbitrary, but we find that many fields
have a similar effect in the evolution of $P_{RL}$.
In Figs.~3(b)-(c) we choose to go back to the initial field value, and
switch the field at the time when $P_{RL}$ in 3(a) is maximum (6.1~ps), 
so that the subsequent oscillation is almost 180 degrees out of phase 
with the oscillation in 3(a).

\begin{figure}
 \vbox to 9cm {\vss\hbox to 6cm
 {\hss\
   {\includegraphics{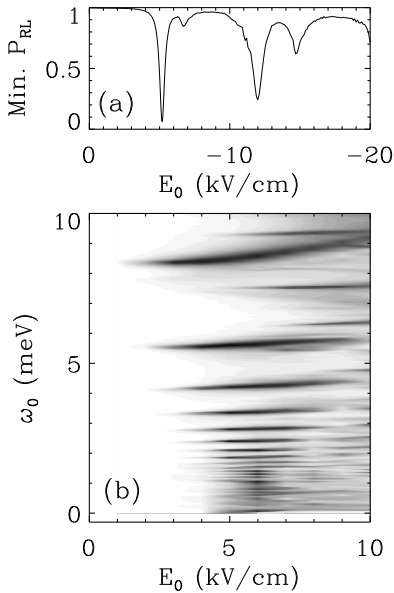}
   }
  \hss}
 }
\caption{
Minimum probability $P_{RL}$ obtained during a time interval
of 9~ps after switching on the external field.
(a) For a constant field of amplitude $E_0$.
(b) For a sinusoidal field of amplitude $E_0$ and frequency $\omega_0$.
Darker areas correspond to lower values of $P_{RL}$, and therefore to
stronger localization.
}
\end{figure}

We saw above that an appropriate constant electric field can localize
to a large extent both electrons {\it in a well of choice} at certain times,
and the localization can be locked by switching the field to another
appropriate value.
We will see next that the nature of the localization is different
when oscillatory fields are applied.
We consider a sinusoidal field of the form $E(t)=E_0 \cos{\omega_0 t}$
(we discuss below the effect of a slow switching-on.)
Fig.~2(b) shows a contour plot of the minimum probability $P_{RL}$
achieved in a simulation interval of $9 \, \mbox{ps}$, as a function
of $E_0$ and $\omega_0$.
Darker areas correspond to lower values of $P_{RL}$, i.e., to stronger
localization.
The most prominent feature in this plot is the existence of a number of
``resonant'' frequencies, which lead to localization for wide
ranges of $E_0$.
The lowest value of $P_{RL}$, 0.027, is obtained for $E_0=5 \, \mbox{kV/cm}$
and $\omega_0=5.6 \, \mbox{meV}$.
Other combinations of $E_0$ and $\omega_0$ also yield values
of $P_{RL}$ below 0.1.
We computed $P_{RL}(t), P_{RR}(t)$ and $P_{LL}(t)$ for the 
frequencies that lead to strong localization and for different 
values of $E_0$,
and in Fig.~4(a) we show the result for the case of strongest
localization.  
Fig.~4(b) shows the corresponding field.
We notice that the localization, or reduction of $P_{RL}$, results
in an increase of {\em both} $P_{RR}$ and $P_{LL}$, as opposed
to the case of a constant field, where only $P_{RR}$ increased
at the expense of $P_{RL}$.
This feature was observed for all the AC fields ($\omega_0$s and $E_0$s)
we looked at.
Physically, the localization with an AC field is such that both electrons 
are (with high probability) together in one of the wells, with roughly 
equal probabilities to be found in either well.
The two electrons are here in quantum states that are both localized 
(to a large extent) and entangled.
We mention that all the states occupied by the two interacting
electrons in our simulations are entangled, in the usual sense that 
they are not factorizable into single-particle states.
We use the label ``localized/entangled'' for the low $P_{RL}$ 
states of Fig.~4(a) to emphasize that, in these states, while each 
individual electron is delocalized (it can be found in either dot 
with roughly 50\% probability), the two electrons are correlated and 
very likely to be found in the same dot.

\begin{figure}
 \vbox to 7cm {\vss\hbox to 6cm
 {\hss\
   {\includegraphics{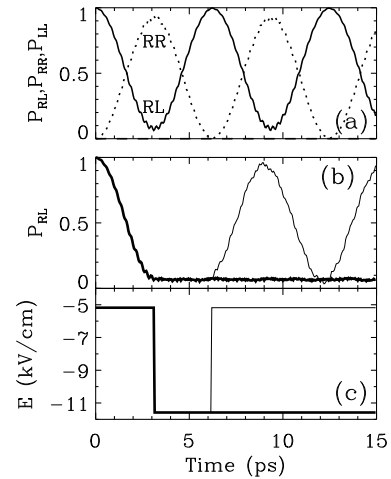}
   }
  \hss}
 }
\caption{
(a) Probabilities that the two electrons are in the different
combinations of wells, for a constant electric field of $E_0=-5.18$~kV/cm,
which produces the highest localization (lowest value of $P_{RL}$ of all
constant fields);
(b) $P_{RL}$ for the two fields shown in (c) (corresponding thick and
thin lines);
(c) Piecewise-constant fields that induce the $P_{RL}(t)$ of (b).
The field values and switching times of the thick-line field are chosen
to produce the lowest $P_{RL}$, and to lock it at its low value;
the thin-line field ads another switching to return to a delocalized
state.  The last switching time can be chosen arbitrarily; in this case
it is such that the oscillation resumes out of phase with the initial one.
}
\end{figure}

A sudden switching-on of the field introduces high frequencies
and therefore population of higher lying excited states,
which results in a rugged evolution of $P_{RL}, P_{RR}$ and $P_{LL}$.
A smoother evolution of these probabilities (Fig.~4(c))
is produced by the field plotted in Fig.~4(d), which is switched on
more slowly.

\begin{figure}
 \vbox to 9cm {\vss\hbox to 6cm
 {\hss\
   {\includegraphics{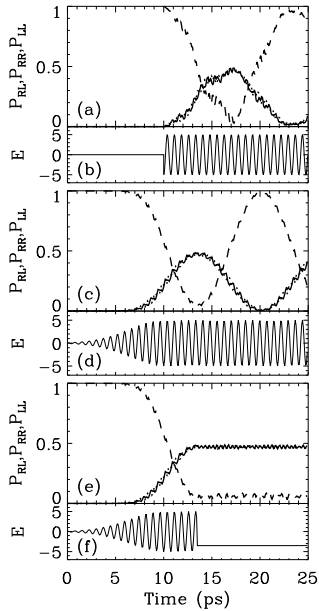}
   }
  \hss}
 }
\caption{
(a) Probabilities that the two electrons are in the different combinations
of wells, for the oscillatory electric field shown in (b).
Dashed line: $P_{RL}$, solid line: $P_{RR}$, dotted line: $P_{LL}$;
(b) $E(t)=E_0 \cos{\omega_0(t-t_0)}$, switched on suddenly at
$t_0=10\,\mbox{ps}$.
$E_0=5\,\mbox{kV/cm}$ and $\omega_0=5.6\,\mbox{meV}$, which gives the
lowest value of $P_{RL}$.
(c) Same as in (a) for the field with slow switching-on shown in (d);
(d) $E(t)=E_0 \exp{[(t-t_0)^2/\lambda^2]} \cos{\omega_0(t-t_0)}$, with
$\lambda=5\,\mbox{ps}$, and $t_0=10\,\mbox{ps}$.
$E_0$ and $\omega_0$ as in (b).
(e) Various probabilities for the field with slow switching-on and
sudden shift to constant value, used to lock in the localized/entangled
state;
(f) Field that produces the probabilities shown in (e).
}
\end{figure}

Once the two electrons are in the localized/entangled
state obtained at $t\approx 13.5\,\mbox{ps}$ in Fig.~4(c), 
they can be forced to stay localized by suddenly shifting the
field to an appropriate constant value.
In Fig.~4(e) we show this effect, along with the electric field
that produces it (shown in (f)).
The control of entangled quantum states in solid state systems
is of great current interest.\cite{los_suk,qui_joh,ima_aws_bur}
Loss and Sukhorukov\cite{los_suk} studied entanglement in coupled 
quantum dots involving the spin degree of freedom.
Here we have identified a possibly complementary method that 
creates localization with entanglement in the spatial wave 
function of two electrons in coupled quantum dots.

In summary, we have found ways to create rapidly and to maintain
localization in a two-electron coupled-quantum dots system with 
uniform electric fields with a simple time dependence.
While a constant electric field creates pure localization of
both electrons in one well, oscillatory fields induce entangled
states that exhibit localization in either dot with roughly
equal probabilities.
This localization-with-entanglement is a purely many-body phenomenon
brought about by the Coulomb interaction between the electrons,
and the method we propose to create it is potentially useful in future 
applications of the physics of entangled states in solid state systems.

We thank Ata\c{c} Imamoglu for useful conversations.
This research was supported in part by QUEST (NSF Grant No. DMR 91-20007).
We made use of Crunch, the parallel computer at UCSB (NSF Grant No. 
CDA 96-01954).

$^{\ast}$ Current address: Departamento de F\'{\i}sica,
FCEN, Universidad de Buenos Aires, (1428) Buenos Aires, 
Argentina.  E-mail: pablot@df.uba.ar.



\begin{references}
\bibitem{localization}
Due to space limitations, a complete list of references cannot 
be given here.
For an overview of the recent literature, see papers that
cite references [1] and [2].
\end{references}
\end{document}